\documentclass[12pt]{article}
\usepackage{graphicx}
\usepackage{amssymb}
\usepackage{amsmath}
\usepackage{color}
\newcommand{\GeV}{\mbox{~GeV}}

\newcommand{\eqn}[1]{&\hspace{-0.6em}#1\hspace{-0.6em}&}
\begin{document}
\baselineskip 0.6cm
%
\begin{titlepage}
\begin{center}

\begin{flushright}
SU-HET-06-2014%
\end{flushright}

\vskip 2cm

{\Large \bf Higgs inflation and Higgs portal dark matter\vspace{3mm}\\ with right-handed neutrinos
}

\vskip 1.2cm

{\large 
Naoyuki Haba, Hiroyuki Ishida, and Ryo Takahashi
}

\vskip 0.4cm

{\em
  Graduate School of Science and Engineering, Shimane University, \\Matsue 690-8504 Japan
}

\vskip 0.2cm


\vskip 2cm

\vskip .5in
\begin{abstract}
We investigate the Higgs inflation and the Higgs portal dark matter with the 
right-handed neutrino. The dark matter and the right-handed neutrino 
in the Higgs inflation play 
important roles in explaining the recent experimental results of the Higgs and 
top masses, and the cosmic microwave background by BICEP2 at the same time. 
This inflation model predicts $805 \GeV \lesssim m_{\rm DM}\lesssim1220~{\rm 
GeV}$ for the DM mass, $1.05 \times10^{14} \GeV\lesssim 
M_R\lesssim2.04 \times10^{14} \GeV$ for the right-handed neutrino mass, and 
$8.42 \lesssim \xi \lesssim 12.4$ for the non-minimal coupling within 
$m_H=125.6\pm0.35 \GeV$ for the Higgs and $M_t=173.34\pm0.76 \GeV$ for 
the top masses.
\end{abstract}
\end{center}
\end{titlepage}
\renewcommand{\thefootnote}{\#\arabic{footnote}} 
\setcounter{footnote}{0}
%
\section{Introduction}
The standard model (SM) has achieved great success in the last few decades. 
In particular, the discovery of the Higgs boson at the LHC experiment \cite{Chatrchyan:2013lba,CMS} was the last key ingredient to finalize the SM.
However, there are several unsolved 
problems in the SM, e.g., explanations of the origin of dark matter (DM), 
the tiny active neutrino masses, and the inflation, etc. Regarding the inflation, 
the BICEP2 experiment has recently reported the non-vanishing tensor-to-scalar 
ratio~\cite{Ade:2014xna}\footnote{Actually, the joint analysis by BICEP, the Keck Array, and Planck concludes that this signal can be explained by dust~\cite{Ade:2015tva}. However, this setup is still attractive from phenomenological points of view. Furthermore, we will show that the main results do not change much even if we take the tensor-to-scalar ratio as $r=0.048$ which is the central value reported in Ref.~\cite{Ade:2015tva}.}:
 \begin{eqnarray}
  r=0.20_{-0.05}^{+0.07}\,.
 \end{eqnarray}
This result has not been confirmed yet, 
but a number of explanations for the result have since been presented. 
One interesting attempt is given in the context of the Higgs 
inflation~\cite{CervantesCota:1995tz}-\cite{Hamada:2013mya} 
(see Refs.~\cite{Nakayama:2014koa}-\cite{Ko:2014eia} for recent discussions in the 
ordinary Higgs inflation and related works after the BICEP2 result). 
In particular, Ref.~\cite{Hamada:2014iga} showed that a Higgs potential with 
$\xi=7$ of the non-minimal Higgs coupling and a suitable plateau for the 
inflation can explain the above BICEP2 result. But the plateau is realized by a
 slightly smaller top mass than the experimental range as $M_t=173.34\pm0.76$ 
GeV~\cite{ATLAS:2014wva}. 
Then, Ref.~\cite{Haba:2014zda} pointed out that the Higgs inflation with the Higgs portal DM 
(see, e.g., Refs.~\cite{Silveira:1985rk}-\cite{Haba:2014sia}) and the right-handed 
neutrino can explain the result with the experimental central values of the Higgs
 and top masses, and $\xi\simeq10.1$. 
In this scenario, DM (a real singlet scalar) and the right-handed neutrino play a crucial role in the realization of a suitable Higgs potential, {\it i.e.} these particles are important for high-energy behavior in the evolution of the Higgs self-coupling $\lambda(\mu)$ 
obeying the renormalization group equation (RGE). In addition, the right-handed 
neutrino can generate the tiny active neutrino mass through the type-I seesaw 
mechanism.

Reference~\cite{Haba:2014zda} found one solution, 
in which the central values were taken for the Higgs and top masses, 
for realistic inflation. 
For the quantum corrections to the coupling constants ($\beta$-functions) in the model, the 
numerical analyses included the next-to-leading-order (NLO) contributions of the SM particles and the leading-order (LO) ones from DM and the right-handed neutrino. In this work, we will analyze the model based on the NLO computations for both the SM and the singlet 
particles. In addition, the latest experimental errors for the Higgs and top 
masses will be taken into account in the calculations. 

As a result, we will 
point out that this inflation model can explain the results of cosmological 
observations within regions for $805 \GeV \lesssim m_{\rm DM}\lesssim1220~{\rm
 GeV}$ for the DM mass, $1.05 \times10^{14} \GeV \lesssim 
M_R\lesssim 2.04 \times10^{14} \GeV$ for the right-handed neutrino mass, and 
$8.42 \lesssim \xi \lesssim 12.4$ for the non-minimal Higgs coupling to the Ricci 
scalar with $m_H=125.6\pm0.35 \GeV$ \cite{Hahn:2014qla} for the Higgs and 
$M_t=173.34\pm0.76 \GeV$ \cite{ATLAS:2014wva} for the top masses. 
A strong correlation between 
the DM and the right-handed neutrino masses will also be shown 
and the allowed region of a DM mass will be checked by future DM detection experiments.

The paper is organized as follows: In section 2, we will explain our model. In 
section 3, we will briefly review the context of the Higgs inflation.
Our numerical results will be given in section 4. Section 5 is devoted to the conclusions. 
We will also present the relevant $\beta$-functions up to 2-loop level 
in the model in the Appendix.

\section{Extension of the Standard Model}
In this letter, we consider the extended SM with a real singlet scalar 
and right-handed neutrinos.
The adequate Lagrangians can be written as
\begin{eqnarray}
\mathcal{L} \eqn{=} 
\mathcal{L}_{\rm SM} + \mathcal{L}_S + \mathcal{L}_N\,,\\
\mathcal{L}_{\rm SM} \eqn{\supset} -\lambda \left( \left|H\right|^2 - \frac{v_{\rm EW}^2}{2} \right)^2\,,\label{Eq:Higgs}\\
\mathcal{L}_S \eqn{=} -\frac{m_S^2}{2} S^2 - \frac{k}{2} \left|H\right|^2 S^2 
- \frac{\lambda_S}{4!} S^4 + \left( {\rm kinetic~ terms} \right)\,,\\
\mathcal{L}_N \eqn{=} - \left( \frac{M_R}{2} \bar{N^c} N + y_N \bar{L} H N + {\rm c.c.} \right)
+ \left( {\rm kinetic~ terms} \right)\,,
\end{eqnarray}
where $\mathcal{L}_{\rm SM}$ is the SM Lagrangian in which the Higgs potential is included as shown in Eq. (\ref{Eq:Higgs}).
$H$ is the Higgs doublet field and $v_{\rm EW}$ is its vacuum expectation value.
$L$, $S$, and $N$ are the lepton doublet, an SM gauge singlet real scalar, and the right-handed neutrino fields, respectively.
The coupling constants $k$, $\lambda_S$, and $y_N$ are 
the portal coupling of $H$ and $S$, the quartic self-coupling of $S$, 
and the neutrino Yukawa coupling constant.
$M_R$ is the right-handed neutrino mass.
We assume that the singlet scalar has odd parity under an additional $Z_2$ symmetry.
Hence, we can have a candidate for DM 
when the mass and the portal coupling $k$ of $S$ are taken to be appropriate values.
The observed tiny neutrino masses can be obtained by the conventional type-I seesaw mechanism\footnote{The right-handed neutrino with the mass $M_R$ generates 
one active neutrino mass. 
Others can be obtained by lighter right-handed neutrinos with smaller Yukawa coupling. 
When the neutrino Yukawa couplings are smaller than the bottom Yukawa coupling, 
the contributions from the neutrino Yukawa couplings to the $\beta$-functions are negligible.
In this work, we consider a case in which the neutrino Yukawa coupling of only one generation of 
the right-handed neutrino is effective in the $\beta$-functions.}. We have fixed the active neutrino mass as $0.1$ eV here.

Here, we show the RGEs to solve:
\begin{eqnarray}
(4 \pi)^2 \frac{dX}{dt} \eqn{=} \beta_X\,,
\end{eqnarray}
where $X$ symbolizes the SM gauge couplings, the top and the neutrino Yukawa couplings, 
and the scalar quartic couplings in our model,  
and we define $t \equiv \ln ( \mu/1 \GeV)$ with the renormalization scale $\mu$.
In this analysis, we divide the energy region into three parts between the $Z$ boson mass scale, 
$M_Z$, and the Planck scale, $M_{\rm pl} = 2.435 \times 10^{18} \GeV$.
Each part can be considered as $M_Z \leq \mu < m_S$, $m_S \leq \mu < M_R$, 
and $M_R \leq \mu \leq M_{\rm pl}$.
The behavior of $\lambda(\mu)$ can be roughly supposed as follows:
At first, the evolution of $k$ becomes small as $k(M_Z)$ becomes small, 
because the $\beta$-function for $k$ is proportional to $k$ itself.
In this case, the evolution of $\lambda (\mu)$ is really close to the SM one.
Second, 
the contribution from the additional scalar pushes up the evolution of $\lambda(\mu)$.
Thus, we can make $\lambda(\mu)$ positive up to $M_{\rm pl}$ 
with sufficient magnitude of the additional scalar contribution.
On the other hand, 
the contribution from the right-haded neutrinos pulls down $\lambda(\mu)$ 
in the region of $M_R \leq \mu \leq M_{\rm pl}$. 
\begin{figure}[t]
\begin{center}
\includegraphics[width=7cm,clip]{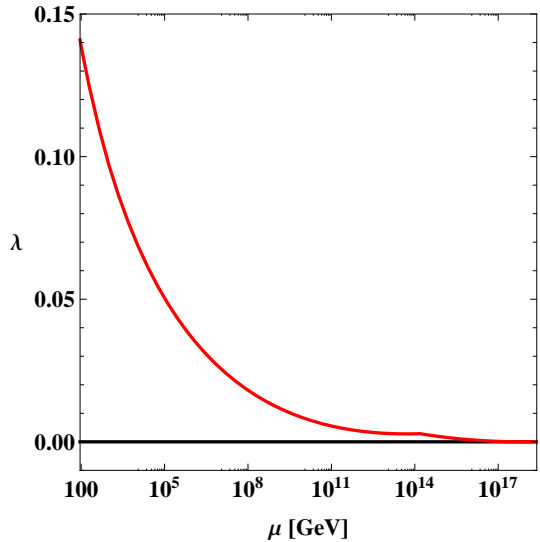}
\caption{A typical behavior of Higgs quartic coupling with central values of the Higgs and top masses.}\label{Fig:running}
\end{center}
\end{figure}
A typical behavior of the running of the Higgs quartic coupling is shown in Fig.\ref{Fig:running}.
As a result, we can have a suitable Higgs potential with a plateau for the inflation 
around $\mathcal{O} (10^{18}) \GeV$~\cite{Haba:2014zda}.
These are the important features for Higgs inflation with a real scalar field 
and right-handed neutrinos\footnote{We assume that $\lambda_S (M_Z)=0.1$ just as a sample point.}.
In our estimation, the free parameters are the values of the right-handed neutrino and DM masses 
and non-minimal coupling $\xi$.

\section{Higgs inflation with singlets}
First, we briefly review the ordinary Higgs inflation \cite{Bezrukov:2007ep} 
with the action in the so-called Jordan frame, 
\begin{eqnarray}
S_J \eqn{\supset} 
\int d^4 x \sqrt{-g} \left( - \frac{M_{\rm pl}^2 +\xi h^2}{2} R + \mathcal{L}_{\rm SM} \right)\,,
\end{eqnarray}
where $\xi$ is the non-minimal coupling of the Higgs to the Ricci scalar $R$, 
$H=(0\,,h)^T / \sqrt{2}$ is given in the unitary gauge, 
and $\mathcal{L}_{\rm SM}$ includes the Higgs potential of Eq. (\ref{Eq:Higgs}).
With the conformal transformation ($\hat{g}_{\mu \nu} \equiv \Omega^2 g_{\mu \nu}$ 
with $\Omega^2 \equiv 1+\xi h^2/M_{\rm pl}^2$) which denotes the transformation 
from the Jordan frame to the Einstein one, 
one can rewrite the action as 
\begin{eqnarray}
S_E \eqn{\supset} 
\int d^4 x \sqrt{-\hat{g}} \left( - \frac{M_{\rm pl}^2}{2} \hat{R} 
+\frac{\partial_\mu \chi \partial^\mu \chi}{2} 
- \frac{\lambda}{4 \Omega (\chi)^4} \left( h (\chi)^2 - v_{\rm EW}^2 \right)^2 \right)\,,
\end{eqnarray}
where $\hat{R}$ is the Ricci scalar in the Einstein frame given by $\hat{g}_{\mu \nu}$, 
and $\chi$ is a canonically renormalized field as
\begin{eqnarray}
\frac{d\chi}{dh} \eqn{=} 
\sqrt{\frac{\Omega^2+6 \xi^2 h^2 / M_{\rm pl}^2}{\Omega^4}}\,.\label{Eq:chi}
\end{eqnarray}
One can calculate the slow-roll parameters as
\begin{eqnarray}
\epsilon \eqn{=} 
\frac{M_{\rm pl}^2}{2} \left( \frac{dU/d \chi}{U} \right)^2\,,
~~
\eta = M_{\rm pl}^2 \frac{d^2U/d \chi^2}{U}\,,
\end{eqnarray}
where $U(\chi)$ is obtained \footnote{In this analysis, we take only the quartic term of the Higgs field into account because the quadratic coupling can be negligible at the inflationary scale. Moreover, in our calculation, we use the $\beta$-functions of coupling constants including the self-coupling of the Higgs up to 2-loop level. This means that our analysis also contains quantum effects, which partially include effects from the loop level potential.} as
\begin{figure}[t]
\begin{center}
\includegraphics[width=8cm,clip]{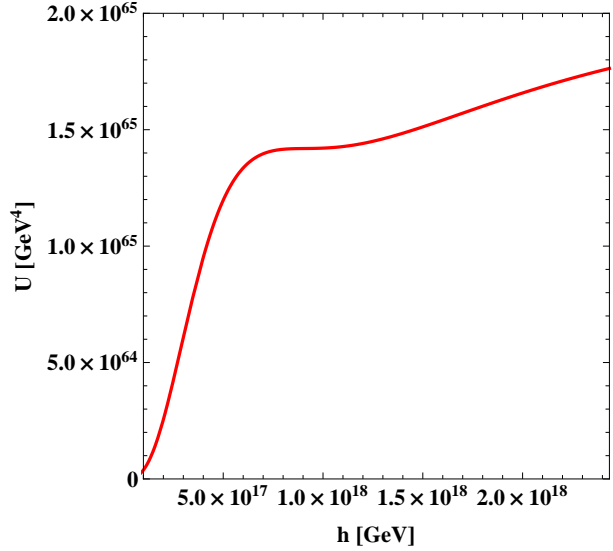}
\caption{A typical behavior of the Higgs potential with central values of the Higgs and top masses.}\label{Fig:Higgs_pot}
\end{center}
\end{figure}
\begin{eqnarray}
U(\chi) \eqn{\equiv} 
\frac{\lambda}{4\Omega(\chi)^4} \left( h (\chi)^2 - v_{\rm EW}^2 \right)^2\,.
\end{eqnarray}
A typical behavior of the Higgs potential $U(\chi)$ is shown in Fig.\ref{Fig:Higgs_pot}.
Using slow-roll parameters, the spectral index and the tensor-to-scalar ratio 
are evaluated as $n_s = 1-6 \epsilon +2 \eta$ and $r=16 \epsilon$, respectively.
Finally, the number of e-foldings is given by 
\begin{eqnarray}
N \eqn{=} 
\int_{h_{\rm end}}^{h_0} \frac{1}{M_{\rm pl}^2} \frac{U}{dU/dh} \left( \frac{d\chi}{dh} \right)^2 dh\,,
\end{eqnarray}
where $h_0$ $(h_{\rm end})$ is the initial (final) value of $h$ corresponding to 
the beginning (end) of the inflation.
At the point $h_{\rm end}$, the slow-roll conditions $(\epsilon\,, |\eta| \ll 1)$ are broken.

The Higgs inflation can be realized even in the SM if the top mass is fine-tuned as $M_t = 171.079 \GeV$ for $m_H=125.6 \GeV$ \cite{Hamada:2013mya,Hamada:2014iga,Fairbairn:2014nxa}. 
With these values, the Higgs potentials have a plateau and 
$r \simeq 0.2$ can be achieved by taking $\xi=7$ \cite{Hamada:2014iga}.
Even though this framework can accomplish a relevant amount of e-foldings, 
the required top mass, $M_t \simeq 171.1 \GeV$, is outside $M_t=173.34 \pm 0.76 \GeV$ \cite{ATLAS:2014wva}.
On the other hand, without the plateau, 
a sufficient amount of e-foldings can be obtained by assuming $\xi \sim \mathcal{O}(10^4)$.
However, this case is plagued with too tiny a tensor-to-scalar ratio on the order of $10^{-3}$, 
which is inconsistent with the recent BICEP2 result.
Consequently, 
we extend the SM with a real scalar and right-handed neutrinos, 
in which the evolution of $\lambda$ (equivalent to the Higgs potential) is changed, 
in order to reproduce the values of cosmological parameters 
within the experimental range $M_t$.

\section{Numerical analysis at 2-loop order}
In this section, we give the results of our numerical analysis.
We solve RGEs\footnote{Here, we consider that this renormalization scale is the same as $h/\Omega$.} at 2-loop level for the $\beta$-functions of the relevant couplings in the model\footnote{There are theoretical uncertainties between the low- and high-energy parameters as discussed in Refs. \cite{Bezrukov:2010jz,Bezrukov:2014bra}. But we assume such uncertainties are small enough and can be neglected.}.
As we discussed above, we analyze within the experimental ranges of the Higgs mass $m_h= 125.6 \pm 0.35 \GeV$ \cite{Hahn:2014qla} and the top mass $M_t=173.34 \pm 0.76 \GeV$ \cite{ATLAS:2014wva}. 

As we mentioned in Sec.2, 
we assume that the additional real scalar is DM.
In the case that the DM mass is greater than the Higgs mass ($m_S > m_h$), 
the portal coupling $k(M_Z)$ is well approximated as \cite{Cline:2013gha,Hamada:2014xka} 
\begin{eqnarray}
\log_{10} k (M_Z) \simeq -3.63+1.04 \log_{10} \left( \frac{m_{\rm DM}} \GeV \right)\,,\label{Eq:DM_CONST}
\end{eqnarray}
where $m_{\rm DM}$ denotes the mass of DM given by $m_{\rm DM}^2 = m_S^2 + k v_{\rm EW}^2/2$.

\begin{figure}[t]
\begin{center}
\includegraphics[width=8cm,clip,angle=-90]{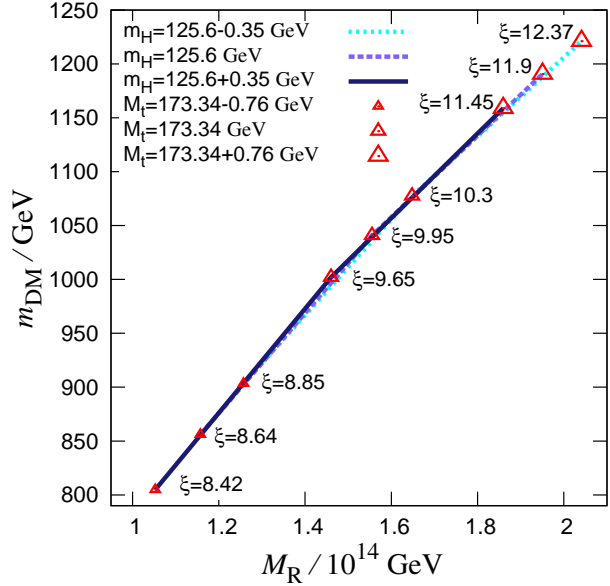}
\end{center}
\caption{The region of $M_R$ and $m_{\rm  DM}$, which reproduces the cosmological parameters of $n_s = 0.9600 \pm 0.0071$ \cite{Ade:2013uln}, $r= 0.20^{+0.07}_{-0.05}$ \cite{Ade:2014xna}, 
and $52.3 \lesssim N \lesssim 59.7$. All points can realize $r=0.2$ within the significant digit.}\label{Fig:MR_MS}
\end{figure}
Our results are shown in Fig.\ref{Fig:MR_MS}.
One can see that the required values of $M_R$, $m_{\rm DM}$, and $\xi$ 
for reproducing suitable cosmological parameters are on a line in the $M_R$-$m_{\rm DM}$ plane.
This is because the form of the Higgs potential is strictly constrained and it should be uniquely 
realized by taking suitable values of $M_R$, $m_{\rm DM}$, and $\xi$ for given $m_H$ and $M_t$.
A lighter (heavier) top (Higgs) mass gives lower bounds on $M_R$, $m_{\rm DM}$, and $\xi$, 
while a heavier (lighter) top (Higgs) mass leads to upper bounds on these parameters.

Finally, we show the explicit range of $M_R$, $m_{\rm DM}$, and $\xi$ for each 
Higgs mass\footnote{There are theoretical uncertainties in matching the high- and 
low-energy parameters~\cite{Bezrukov:2010jz}. For example, they change by about 0.5
 GeV for the 126 GeV Higgs mass. This leads to uncertainties of about $\pm30$ 
GeV for the DM mass, $\pm 5\times10^{13}$ GeV for the right-handed neutrino mass, and
 $\pm0.6$ for the non-minimal coupling in our model.}: 
\begin{itemize}
\item $m_H=125.6-0.35 \GeV$
\begin{eqnarray}
\eqn{}\makebox[-10mm][r]{}
1.26 \times 10^{14} \GeV \lesssim M_R \lesssim 2.04 \times 10^{14} \GeV\,,~~~
903 \GeV \lesssim m_{\rm DM} \lesssim 1221 \GeV\,, \notag \\
\eqn{}\makebox[-10mm][r]{}
8.85 \lesssim \xi \lesssim 12.37\,, \label{14}
\end{eqnarray}
\item $m_H=125.6 \GeV$
\begin{eqnarray}
\eqn{}\makebox[-10mm][r]{}
1.16 \times 10^{14} \GeV \lesssim M_R \lesssim 1.95 \times 10^{14} \GeV\,,~~~
856 \GeV \lesssim m_{\rm DM} \lesssim 1191 \GeV\,,\notag\\
\eqn{}\makebox[-10mm][r]{}
8.64 \lesssim \xi \lesssim 11.9\,,
\end{eqnarray}
\item $m_H=125.6+0.35 \GeV$
\begin{eqnarray}
\eqn{}\makebox[-10mm][r]{}
1.05 \times 10^{14} \GeV \lesssim M_R \lesssim 1.86 \times 10^{14} \GeV\,,~~~
805 \GeV \lesssim m_{\rm DM} \lesssim 1159 \GeV\,,\notag\\
\eqn{}\makebox[-10mm][r]{}
8.42 \lesssim \xi \lesssim 11.45\,. \label{16}
\end{eqnarray}
\end{itemize}

$k(M_Z)$  is determined by a condition that the singlet scalar gives a 
sufficient amount of the relic density of the DM\footnote{In this work, we have neglected the non-thermal production of DM via portal interaction due to the smallness of the cross section. Therefore, we can estimate the total amount of DM abundance by thermal production.}, 
{\it i.e.} $\Omega_{\rm DM} \bar{h}^2 = 0.119$ (see, {\it e.g.}, Refs. \cite{HKT,Cline:2013gha}) 
where $\Omega_{\rm DM}$ is the density parameter of the DM and $\bar{h}$ is the Hubble constant.
From this condition, 
$k(M_Z)$ is in the range $0.25 \lesssim k(M_Z) \lesssim 0.38$.
The region is consistent with the current experimental constraints so far, 
and future experiments, {\it e.g.}, XENON1T or LUX for the direct detection, 
or the combined analysis of the Fermi+CTA+Planck observations might make them clear \cite{Cline:2013gha}.
Finally, we also comment on the validity of our $\beta$-functions. 
Since the value of $\xi$ is around $10$ as in the above results 
and it is much smaller than the previous work \cite{Bezrukov:2009db}, 
we can utilize our $\beta$-functions up to the inflation scale.

\subsection{After the joint analysis of BICEP, Keck Array and Planck}
The joint analysis of BICEP, the Keck Array and Planck showed only an upper bound on the tensor-to-scalar ratio~\cite{Ade:2015tva}\footnote{The latest independent result from the Planck Collaboration gives a consistent result~\cite{Planck:2015xua}.} of 
\begin{eqnarray}
r<0.12\,.
\end{eqnarray}

\begin{figure}[t]
\begin{center}
\includegraphics[width=8cm,clip,angle=-90]{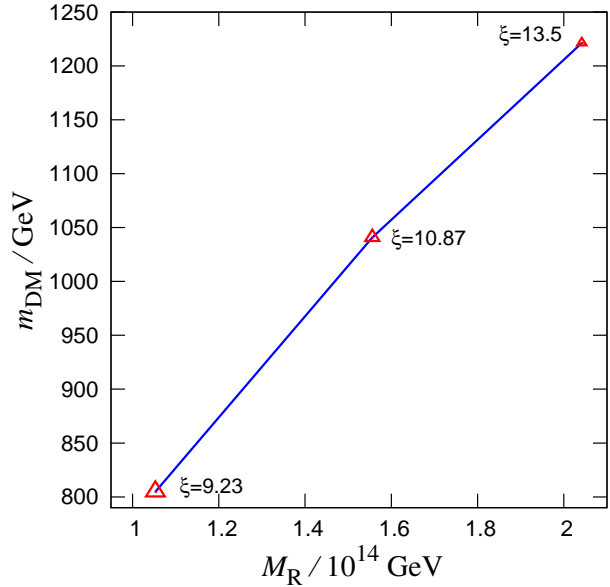}
\end{center}
\caption{The region of $M_R$ and $m_{\rm  DM}$ respecting $r= 0.048^{+0.035}_{-0.032}$ \cite{Ade:2015tva}. We show only three points here and the meaning of each point is same as in Fig.\ref{Fig:MR_MS}. }\label{Fig:MR_MS_0.05}
\end{figure}
We show our result respecting $r= 0.048^{+0.035}_{-0.032}$ in Fig.\ref{Fig:MR_MS_0.05}.
One can easily see that the mass scales of right-handed neutrino and DM 
do not drastically change compared with the $r=0.2$ case.
Therefore, the future direct-detection experiments of DM can reach even if the mass range of DM shown in Fig.\ref{Fig:MR_MS_0.05}. 
The difference appears in the magnitude of non-minimal coupling, $\xi$.
Its magnitude should be slightly larger in the range $9.23 \lesssim \xi \lesssim 13.5$.

\section{Conclusions}
We have investigated the Higgs inflation model with the Higgs portal DM and the 
right-handed neutrino by the use of $\beta$-functions up to 2-loop level. 
In addition, the latest experimental errors of the top and Higgs masses have been taken into account in the calculations. 
As a result, we pointed out that this inflation model can explain the results of cosmological observations $n_s = 0.9600 \pm 0.0071$, $r= 0.20^{+0.07}_{-0.06}$, and $52.3 \lesssim N \lesssim 59.7$ within regions of $805 \GeV \lesssim m_{\rm DM} \lesssim 1220 \GeV$ for the DM mass, $1.05 \times10^{14} \GeV \lesssim M_R \lesssim 2.04 \times10^{14} \GeV$ for 
the right-handed neutrino mass, and $8.42 \lesssim \xi \lesssim 12.4$ for the 
non-minimal Higgs coupling to the Ricci scalar with $m_H=125.6\pm0.35 \GeV$ for the Higgs and $M_t=173.34\pm0.76 \GeV$ for the top masses. 
Furthermore, we have shown that the scales of the right-handed neutrino and DM are not significantly changed even though the tensor-to-scalar ratio decreases as $r<0.12$.
On the other hand, the non-minimal coupling should be slightly larger than the $r=0.2$ case as $9.23 \lesssim \xi \lesssim 13.5$.
There is a strong correlation between the DM and the right-handed neutrino masses 
because the form of the Higgs potential is strictly constrained; it should be uniquely 
realized by taking suitable values of $M_R$, $m_{\rm DM}$, and $\xi$ for given $m_H$ and $M_t$.
The DM mass region in our analysis will be confirmed by future DM detections.

\subsection*{Acknowledgements}

This work is partially supported by a Scientific Grant by the Ministry of Education 
and Science, No. 24540272. The work of R.T. is supported by Research Fellowships
 of the Japan Society for the Promotion of Science for Young Scientists.

\appendix
\section*{Appendix}
\subsection*{Renormalization group equations at 2-loop level}
In this section, we give the $\beta$-functions of the SM gauge couplings, top 
and neutrino Yukawa couplings, and the scalar quartic couplings in our model at 2-loop level\footnote{When one includes the effects of non-minimal coupling in the $\beta$-functions, the desired values of the DM mass, the right-handed neutrino mass, and the non-minimal coupling for the successful Higgs inflation increase a few \% from those values given in Eqs.~(\ref{14})-(\ref{16}).}.
First, 
\begin{eqnarray}
\beta_{g'} \eqn{=} 
\frac{41}{6} g'^3 
+ \frac{1}{16 \pi^2} g'^3 
\left[ \frac{199}{18} g'^2 + \frac{9}{2} g^2 + \frac{44}{3} g_s^2 - \frac{17}{6} y_t^2 
- \frac{1}{2} y_N^2 \right]\,,\\
\beta_{g} \eqn{=} 
- \frac{19}{6} g^3 
+ \frac{1}{16 \pi^2} g^3 \left[ \frac{3}{2} g'^2 + \frac{35}{6} g^2 + 12 g_s^2 - \frac{3}{2} y_t^2 
- \frac{1}{2} y_N^2  \right]\,,\\
\beta_{g_s} \eqn{=} 
-7 g_s^3 
+ \frac{1}{16 \pi^2} g_s^3 
\left[ \frac{11}{6} g'^2 + \frac{9}{2} g^2 -26 g_s^2 -2 y_t^2 \right]\,,
\end{eqnarray}
are for the ${\rm U}(1)_Y$, ${\rm SU}(2)_L$, and ${\rm SU}(3)_c$ gauge couplings. 

Next,
\begin{eqnarray}
\beta_{y_t} \eqn{=} 
y_t \left[ \frac{9}{2} y_t^2 + y_N^2 - \left( \frac{17}{12} g'^2 + \frac{9}{4} g^2 +8 g_s^2 \right) \right]\notag\\
\eqn{} 
+ \frac{1}{16 \pi^2} y_t 
\bigg[ 
-12 y_t^4 + 6 \lambda^2 - 12 \lambda y_t^2 
+ \frac{131}{16} g'^2 y_t^2 + \frac{225}{16} g^2 y_t^2 + 36 g_s^2 y_t^2\notag\\
\eqn{} 
+ \frac{1187}{216} g'^4 - \frac{23}{4} g^4 - 108 g_s^4 - \frac{3}{4} g'^2  g^2 
+ 9 g^2 g_s^2 + \frac{19}{9} g'^2 g_s^2 + \frac{1}{4} k^2 \notag\\
\eqn{}+ \frac{5}{8} g'^2 y_N^2 + \frac{15}{8} g^2 y_N^2 
- \frac{9}{4} y_N^4 - \frac{9}{4} y_N^2 y_t^2
\bigg]\,,\\
\beta_{y_N} \eqn{=} 
y_N \left[ - \frac{3}{4} g'^2 - \frac{9}{4} g^2 + 3 y_t^2 + \frac{5}{2} y_N^2 \right]\notag\\
\eqn{}
+ \frac{1}{16 \pi^2} y_N 
\bigg[
-3 y_N^4 + 6 \lambda^2 -12 \lambda y_N^2 + \frac{103}{16} g'^2 y_N^2 
+ \frac{165}{16} g^2 y_N^2 + \frac{35}{24} g'^4 - \frac{9}{4} g'^2 g^2 - \frac{23}{4} g^4\notag\\
\eqn{}
+ \frac{1}{4} k^2 + \frac{85}{24} g'^2 y_t^2 + \frac{45}{8} g^2 y_t^2 
+ 20 g_s^2 y_t^2 - \frac{27}{4} y_t^2 y_N^2 - \frac{27}{4} y_t^4
\bigg]
\,,
\end{eqnarray}
are for the top and neutrino Yukawa couplings. 

Lastly, 
\begin{eqnarray}
\beta_\lambda \eqn{=} 
24 \lambda^2 - 2 ( 3 y_t^4 + y_N^4 ) + 4 \lambda \left( 3 y_t^2 + y_N^2\right)
-3 \lambda \left( g'^2 + 3 g^2 \right) 
+ \frac{3}{8} \left[ 2 g^4 + ( g'^2 + g^2 )^2 \right] + \frac{1}{2} k^2\notag\\
\eqn{}
+\frac{1}{16 \pi^2} \bigg[ 
-312 \lambda^3+36 \lambda^2 \left ( g'^2 + 3 g^2 \right) - 
\lambda \left( -\frac{629}{24} g'^4 -\frac{39}{4} g'^2 g^2 + \frac{73}{8} g^4 \right)\notag\\
\eqn{}
+\frac{305}{16} g^6 - \frac{289}{48} g'^2 g^4 - \frac{559}{48} g'^4 g^2 - \frac{379}{48} g'^6
-32 g_s^2 y_t^4 -\frac{8}{3} g'^2 y_t^4 - \frac{9}{4} g^4 y_t^2\notag\\
\eqn{}
+\lambda y_t^2 \left( \frac{85}{6} g'^2 + \frac{45}{2} g^2 + 80 g_s^2 \right) 
+ g'^2 y_t^2 \left( -\frac{19}{4} g'^2 + \frac{21}{2} g^2 \right)\notag\\
\eqn{}
-144 \lambda^2 y_t^2 - 3 \lambda y_t^4 + 30 y_t^6 -5 k^2 \lambda -2 k^3 
- 48 \lambda^2 y_N^2 - \frac{1}{4} g'^4 y_N^2- \frac{1}{2} g'^2 g^2 y_N^2 - \frac{3}{4} g^4 y_N^2 \notag\\
\eqn{}
+ \lambda  y_N^2  \left( \frac{5}{2} g'^2 + \frac{15}{2} g^2 \right) 
- \lambda y_N^4 +10 y_N^6
\bigg] 
\,,\\
\beta_k \eqn{=} 
k \left[ 
4 k +12 \lambda + \lambda_S +6 y_t^2 + 2 y_N^2 - \frac{3}{2} \left( g'^2 + 3g^2  \right) \right]\notag\\
\eqn{}
+ \frac{1}{16 \pi^2} 
\bigg[
- \frac{21}{2} k^3 - 6 k^2 \left( 12 \lambda + \lambda_S \right) 
-5 k \left( 12 \lambda^2 + \frac{1}{6} \lambda_S^2 \right) 
+ k^2 \left( g'^2 + 3g^2 \right) \notag\\
\eqn{}
+ \frac{1}{8} k \left( \frac{557}{6} g'^4 + 15 g'^2 g^2 - \frac{145}{2} g^4 \right) 
+ 5 k y_t^2 \left( \frac{17}{12} g'^2 + \frac{9}{4} g^2 + 8 g_s^2 \right) 
+ 24 k \lambda \left( g'^2 + 3 g^2 \right) \notag\\
\eqn{}
-12 k^2 y_t^2 
- \frac{27}{2} k y_t^4 - 72 k \lambda y_t^2 
+ \frac{5}{4} k y_N^2 \left( g'^2 + 3g^2 \right) 
-4 k^2 y_N^2 - \frac{9}{2} k y_N^4 -24 k \lambda y_N^2
\bigg]
\,,\\
\beta_{\lambda_S} \eqn{=} 
3 \lambda_S^2+12 k^2+
\frac{1}{16 \pi^2} \left[ 
-\frac{17}{3} \lambda_S^3 - 20 k^2 \lambda_S - 48 k^3 + 
24 k^2 \left( (g'^2 + 3 g^2) - ( 3 y_t^2 + y_N^2) \right) \right]\,,
\end{eqnarray}
are for the SM Higgs quartic, the portal between the SM Higgs and the singlet scalar, 
and the singlet scalar quartic couplings.


\end{document}